\documentclass{emulateapj}

\def\ms{m~s$^{-1}$}
\def\ks{km~s$^{-1}$}
\def\msini{$M_P\sin{i}$}

\def\logg{$\log{g}$}
\def\vsini{$V_{\rm rot}\sin{i}$}
\def\teff{$T_{\rm eff}$}
\def\msun{$M_{\odot}$}
\def\mjup{$M_{\rm Jup}$}
\def\rsun{$R_{\odot}$}
\def\lsun{$L_{\odot}$}
\def\chisq{$\sqrt{\chi^2_\nu}$}

\def\feh{[Fe/H]}
\def\rphk{$R^\prime_{HK}$}

\def\NBODYpAb{452.8} 
\def\NBODYpeAb{$^{+2.1}_{-4.5}$}
\def\NBODYtAb{2454762} 
\def\NBODYteAb{$^{+67.3}_{-172.3}$}
\def\NBODYeAb{0.09} 
\def\NBODYeeAb{$^{+0.14}_{-0.06}$}
\def\NBODYkAb{40.0} 
\def\NBODYkeAb{$^{+4.9}_{-7.8}$}
\def\NBODYomAb{9.2} 
\def\NBODYomeAb{$^{+277.9}_{-165.4}$}
\def\NBODYmsiniAb{1.99} 
\def\NBODYmsinieAb{$^{+0.26}_{-0.38}$}
\def\NBODYarelAb{1.333} 
\def\NBODYareleAb{$^{+0.004}_{-0.009}$}

\def\NBODYpAc{883.0} 
\def\NBODYpeAc{$^{+32}_{-14}$}
\def\NBODYtAc{2454930} 
\def\NBODYteAc{$^{+209.9}_{-96.5}$}
\def\NBODYeAc{0.29} 
\def\NBODYeeAc{$^{+0.16}_{-0.09}$}
\def\NBODYkAc{14.5} 
\def\NBODYkeAc{$^{+7.5}_{-3.6}$}
\def\NBODYomAc{220.5} 
\def\NBODYomeAc{$^{+182.2}_{-320.9}$}
\def\NBODYmsiniAc{0.86} 
\def\NBODYmsinieAc{$^{+0.35}_{-0.22}$}
\def\NBODYarelAc{2.08} 
\def\NBODYareleAc{$^{+0.05}_{-0.02}$}

\def\NBODYjA{9.9} 
\def\NBODYjeA{$^{+2.9}_{-1.2}$}

\def\NBODYpBb{613.8} 
\def\NBODYpeBb{$^{+1.3}_{-1.4}$}
\def\NBODYtBb{2454900} 
\def\NBODYteBb{$^{+235.5}_{-89.3}$}
\def\NBODYeBb{0.04} 
\def\NBODYeeBb{$^{+0.04}_{-0.02}$}
\def\NBODYkBb{34.5} 
\def\NBODYkeBb{$^{+2.7}_{-1.5}$}
\def\NBODYomBb{288.0} 
\def\NBODYomeBb{$^{+47.0}_{-111.9}$}
\def\NBODYmsiniBb{1.85} 
\def\NBODYmsinieBb{$^{+0.14}_{-0.08}$}
\def\NBODYarelBb{1.601} 
\def\NBODYareleBb{$^{0.002}_{-0.002}$}
 
\def\NBODYpBc{825.0} 
\def\NBODYpeBc{$^{+5.1}_{-3.1}$}
\def\NBODYtBc{2455000} 
\def\NBODYteBc{$^{+51.1}_{-54.6}$}
\def\NBODYeBc{0.181} 
\def\NBODYeeBc{$^{+0.024}_{-0.017}$}
\def\NBODYkBc{15.42} 
\def\NBODYkeBc{$^{+2.19}_{-1.04}$}
\def\NBODYomBc{182.6} 
\def\NBODYomeBc{$^{+67.7}_{-57.1}$}
\def\NBODYmsiniBc{0.90} 
\def\NBODYmsinieBc{$^{+0.12}_{-0.06}$}
\def\NBODYarelBc{1.95} 
\def\NBODYareleBc{$^{+0.008}_{-0.005}$}

\def\NBODYjB{8.23} 
\def\NBODYjeB{$^{+0.38}_{-0.88}$}

\def\starA{HD\,90043}
\def\pAb{455.2}
\def\peAb{3.2}
\def\pAc{910}
\def\peAc{21}
\def\tpAb{2454758}
\def\tpeAb{30}
\def\tpAc{2454941}
\def\tpeAc{30}
\def\eAb{0.184}
\def\eeAb{0.029}
\def\eAc{0.412}
\def\eeAc{0.064}
\def\omAb{227}
\def\omeAb{20}
\def\omAc{172.0}
\def\omeAc{9}
\def\kAb{33.2}
\def\keAb{1.6}
\def\kAc{23.5}
\def\keAc{2.9}
\def\msiniAb{1.6}
\def\msinieAb{0.2}
\def\msiniAc{1.4}
\def\msinieAc{0.2}
\def\arelAb{1.41}
\def\areleAb{0.03}
\def\arelAc{2.22}
\def\areleAc{0.06}
\def\rmslA{7.7}
\def\rmskA{4.8}
\def\chisA{1.14}
\def\nobslA{50}
\def\nobskA{24}
\def\mstarA{1.54}
\def\mstareA{0.08}
\def\bvA{0.92}
\def\bveA{0.01}
\def\vmagA{6.61}
\def\vmageA{0.04}
\def\mvA{2.17}
\def\mveA{0.06}
\def\vsiniA{2.77}
\def\ageA{2.7}
\def\ageeA{0.4}
\def\rstarA{4.9}
\def\rstareA{0.08}
\def\lstarA{14.6}
\def\lstareA{0.1}
\def\teffA{5098}
\def\loggA{3.5}
\def\feA{-0.03}
\def\dA{74.8}
\def\deA{4.9}
\def\starB{HD\,200964}
\def\pBb{630.6}
\def\peBb{9.3}
\def\pBc{829}
\def\peBc{21}
\def\tpBb{2454916}
\def\tpeBb{30}
\def\tpBc{2455029}
\def\tpeBc{130}
\def\eBb{0.111}
\def\eeBb{0.030}
\def\eBc{0.113}
\def\eeBc{0.076}
\def\omBb{223}
\def\omeBb{20}
\def\omBc{301}
\def\omeBc{50}
\def\kBb{35.2}
\def\keBb{2.7}
\def\kBc{22.1}
\def\keBc{2.3}
\def\msiniBb{1.9}
\def\msinieBb{0.2}
\def\msiniBc{1.3}
\def\msinieBc{0.2}
\def\arelBb{1.71}
\def\areleBb{0.04}
\def\arelBc{2.03}
\def\areleBc{0.06}
\def\rmslB{7.6}
\def\rmskB{5.3}
\def\chisB{1.15}
\def\nobslB{61}
\def\nobskB{35}
\def\mstarB{1.44}
\def\mstareB{0.09}
\def\bvB{0.880}
\def\bveB{0.009}
\def\vmagB{6.64}
\def\vmageB{0.04}
\def\mvB{2.35}
\def\mveB{0.07}
\def\vsiniB{2.28}
\def\ageB{3.0}
\def\ageeB{0.6}
\def\rstarB{4.3}
\def\rstareB{0.09}
\def\lstarB{11.6}
\def\lstareB{0.4}
\def\teffB{5164}
\def\loggB{3.6}
\def\feB{-0.15}
\def\dB{68.4}
\def\deB{4.8}

\def\starA{24~Sex}
\def\starB{HD\,200964}

\begin{document}
\title{Retired A Stars and Their Companions VI. \\A Pair of Interacting
  Exoplanet Pairs Around the Subgiants 24~Sextanis and HD\,200964$^1$} 

\author{ John Asher Johnson\altaffilmark{2},
Matthew Payne\altaffilmark{3},
Andrew W. Howard\altaffilmark{4},
Kelsey I. Clubb\altaffilmark{5},
Eric B. Ford\altaffilmark{3},
Brendan P. Bowler\altaffilmark{6},
Gregory W. Henry\altaffilmark{7},
Debra A. Fischer\altaffilmark{8},
Geoffrey W. Marcy\altaffilmark{4},
John M. Brewer\altaffilmark{8},
Christian Schwab\altaffilmark{9,8},
Sabine Reffert\altaffilmark{9},
Thomas B. Lowe\altaffilmark{10}
}

\email{johnjohn@astro.caltech.edu}

\altaffiltext{1}{Based on observations obtained at the Lick
  Observatory, which is operated by the University of California, and
  W. M. Keck Observatory, which is operated jointly by the University
 of California and the California Institute of Technology}
\altaffiltext{2}{California Institute of Technology,
  Department of Astrophysics, MC 249-17, Pasadena, CA 91125} 
\altaffiltext{3}{Department of Astronomy, University of Florida, 211
  Bryant Space Science Center, PO Box 112055, Gainesville, FL
  32611-2055, USA}  
\altaffiltext{4}{Department of Astronomy, University of California,
  Mail Code 3411, Berkeley, CA 94720}  
\altaffiltext{5}{Department of Physics \& Astronomy, San Francisco
  State University, San Francisco, CA 94132}  
\altaffiltext{6}{Institute for Astronomy, University of Hawai'i,
  Honolulu, HI 96822}
\altaffiltext{7}{Center of Excellence in Information Systems, Tennessee
  State University, 3500 John A. Merritt Blvd., Box 9501, Nashville, TN 37209}
\altaffiltext{8}{Department of Astronomy, Yale University, New Haven, CT 06511}
\altaffiltext{9}{ZAH-Landessternwarte, K\"onigstuhl 12, 69117
  Heidelberg, Germany}
\altaffiltext{10}{UCO/Lick Observatory, Santa Cruz, CA 95064, USA}

\begin{abstract}
We report radial velocity measurements of the G-type subgiants
24~Sextanis (=HD\,90043) and HD\,200964. Both are massive, evolved
stars that exhibit periodic variations due to the presence of
a pair of Jovian planets. Photometric monitoring with the T12 0.80~m
APT at Fairborn Observatory demonstrates both stars to be constant in
brightness to $\leq 0.002$~mag, thus strengthening the planetary
interpretation of the radial velocity variations.  \starA\,b,\,c have
orbital periods of \NBODYpAb~days and \NBODYpAc~days, corresponding to
semimajor axes \NBODYarelAb~AU 
and \NBODYarelAc~AU, and minimum masses \NBODYmsiniAb~\mjup\ and
\NBODYmsiniAc~\mjup, assuming a stellar mass $M_\star =
$~\mstarA~\msun. \starB\,b,\,c have orbital   
periods of \NBODYpBb~days and \NBODYpBc~days, corresponding to semimajor axes
\NBODYarelBb~AU and \NBODYarelBc~AU, and minimum masses \NBODYmsiniAb~\mjup\ and
\NBODYmsiniBc~\mjup, assuming $M_\star = $~\mstarB~\msun. We also
carry out dynamical simulations to properly 
account for gravitational interactions between the planets. 
Most, if not all, of the dynamically stable solutions include crossing
orbits, suggesting that each system is
locked in a mean motion resonance that prevents close encounters and
provides long-term stability.  The planets in the \starA\ system likely
have a period ratio 
near 2:1, while the \starB\ system is even more tightly
packed with a period ratio close to 4:3. However, we caution that
further radial velocity observations and more detailed dynamical
modelling will be required to provide definitive and unique orbital
solutions for both cases, and to determine whether the two systems are
truly resonant.  
\end{abstract}

\keywords{techniques: radial velocities---planetary systems:
  formation---stars: individual (\starA, \starB)}

\section{Introduction}

The giant planets thus far discovered around other stars exhibit a wide
variety of orbital characteristics that are very different from the
properties of the 
planets in our Solar System. For example, exoplanets rarely reside in
circular orbits and many of them have semimajor axes 1 to 2 orders of
magnitude smaller than those of Jupiter and Saturn. However, at least
one characteristic of 
exoplanets appears to be shared with the constituents of the Solar
System: planets often come in bunches. As Wright et al. 2009 recently
showed, 28\% of apparently singleton exoplanets are later discovered
to reside in systems containing two or more components. As the precision
and time baselines of Doppler surveys increase, and as more planets
are discovered from wide-field transit surveys \citep{bakos09b,wasp1},
direct imaging \citep{marois08,kalas08}, and microlensing
\citep{gaudi08,gould10}, the currently measured multiplicity rate will
likely prove to be a lower bound.

Multiplanet systems are manifested in radial velocity
(RV) time series 
as either a single-planet Keplerian motion superimposed atop a partial
longer-period orbit, or as multiple, time-resolved orbits
\citep[e.g.][]{wright09}. Those in 
the former category will gradually come into focus as the time
baselines of Doppler surveys lengthen, and these ``trend'' systems are
becoming recognized as promising direct-imaging targets. The latter
category, with their well-characterized orbits, are extremely valuable
for understanding the origins of planets and the evolution of their
orbital architectures \citep{ford06r}. 

Observed deviations from pure Keplerian motions reveal 
gravitational interactions among planets that serve as fossil records
of past close encounters and/or convergent orbital migration
\citep{malhotra02,wu02}. A prime
example is the system of planets orbiting
$\upsilon$~Andromedae \citep{butler99}. \citet{ford05b} demonstrated through
dynamical modeling that the current orbital configuration shows
evidence of a violent planet-planet scattering event in the distant
past. Similarly, the two Jovian planets in the Gl\,876 system
currently reside in a mean-motion resonance (MMR) that may be a
reflection of differential migration  after the planets'
formation \citep{marcy01,lee04,laughlin05}. Thus, the discovery of
exoplanets in MMRs is strong support 
for the inward orbital migration that is often invoked to explain the
common existence of giant planets well within the canonical
``snow line.''

Additionally, gravitational interactions among resonant planets
can also place  constraints on both the system inclination with
respect to the sky, as well as mutual inclinations between the
planets, and thereby remove the $\sin{i}$ ambiguity and provide absolute
measurements of the planet masses \citep{rivera05,correia10}. Interactions
observed in certain types of
multiplanet systems can reveal the interior structures of gas giant
planets in vivid detail. In the dramatic case of the system 
of planets around HAT-P-13, the inner planet transits its 
host star and experiences additional gravitational perturbations from
an outer planet near 1~AU \citep{bakos09,winn10}. Depending on the
inclinations of the planets in the system, 
precise follow-up measurements may provide estimates of the tidal Love
number and Q value of the inner planet to a higher precision than is
possible for Jupiter \citep{batygin09,mardling10}.  

We are conducting a Doppler survey of intermediate-mass subgiant stars
at the Lick and Keck Observatories with the goal of understanding the
influence stellar mass on the physical properties,
orbital architectures and multiplicity rates of planetary systems. Our
survey has resulted in the discovery of 14 new singleton exoplanets
\citep{johnson06, 
  johnson07,johnson08a, peek09,bowler10,johnson10b}. 
In this contribution we announce the discovery of two pairs of
Jovian planets orbiting the subgiants 24~Sextanis
($=$HD\,90043) and HD\,200964. 

\section{Spectroscopic Observations}

We began observations of \starA\ and \starB\ at Lick Observatory in
2004--2005 as part of our Doppler survey of intermediate-mass
subgiants. Details of the survey, including target selection and
observing strategy are given in \citet{johnson06b}, \citet{peek09} and
\citet{bowler10}. In 2007 we expanded our survey of subgiants at Keck
Observatory \citep{johnson10b} and  we added \starA\ and \starB\ to
our Keck target list for additional monitoring.  

At Lick Observatory, the Shane 3\,m and 0.6\,m 
Coude Auxiliary Telescopes (CAT) feed the Hamilton spectrometer
\citep{vogt87}, and observations at Keck Observatory were
obtained using the HIRES spectrometer \citep{vogt94}.  
Doppler shifts are measured from each observation
using the iodine cell method described by \citet{butler96} \citep[see
  also][]{marcy92b}. A temperature--controlled Pyrex cell containing
gaseous iodine is placed at the entrance slit of the
spectrometer. The dense set of narrow molecular lines imprinted on
each stellar spectrum from 5000 to 6000~\AA\ provides a
robust wavelength scale for each observation, as well as information
about the shape of the spectrometer's instrumental response
\citep{valenti95}.  

At Lick, typical exposure times of 60 minutes on the CAT and 5 minutes
on the 3~m yield a signal-to-noise ratio (SNR) of $\approx$120 at
the center of the iodine region ($\lambda = 5500$~\AA), providing a
velocity precision of 4.0--5.0~\ms. At Keck, 
typical spectra have SNR~$\approx 180$ at 5500~\AA, resulting in a
velocity precision of 1.5--2.0~\ms. 

In addition to the internal, photon-limited uncertainties, the RV
measurements include an additional noise term due to stellar 
``jitter''---velocity noise in excess of internal errors due to
astrophysical sources such as pulsation and rotational modulation of
surface features \citep[][]{saar98, wright05}. We therefore adopt a
jitter value of 5~\ms\ for our subgiants based on the estimate of
\citet{johnson10b}. This jitter term is added in quadrature to the
internal errors before determining the Keplerian orbital
solutions. For the dynamical analysis in \S~\ref{sec:dynamical} we
allow the jitter to vary as a free parameter in the fitting process.

\section{Stellar Properties}
\label{stellar}

\begin{deluxetable}{lll}[!t]
\tablecaption{Stellar Parameters\label{tab:stars}}
\tablewidth{0pt}
\tablehead{
  \colhead{Parameter} & 
  \colhead{\starA\tablenotemark{a}}    &
  \colhead{\starB}    
}
\startdata
V               & \vmagA~(\vmageA)   & \vmagB~(\vmageB)  \\
$M_V$           & \mvA~(\mveA)   & \mvB~(\mveB)    \\
B-V             & \bvA~(\bveA)    & \bvB~(\bveB)    \\
Distance (pc)   & \dA~(\deA)        & \dB~(\deB)          \\
${\rm [Fe/H]}$  & \feA~(0.04)        & \feB~(0.04)        \\
$T_{eff}$~(K)   & \teffA~(44)        & \teffB~(44)   \\
\vsini~(\ks)    & \vsiniA~(0.5)      & \vsiniB~(0.5)      \\
$\log{g}$       & \loggA~(0.1)       & \loggB~(0.1)   \\
$M_{*}$~(\msun) & \mstarA~(\mstareA) & \mstarB~(\mstareB) \\
$R_{*}$~(\rsun) & \rstarA~(\rstareA) & \rstarB~(\rstareB) \\
$L_{*}$~(\lsun) & \lstarA~(\lstareA) & \lstarB~(\lstareB) \\
Age~(Gyr)       & \ageA~(\ageeA)     & \ageB~(\ageeB)     \\
$\log R'_{HK}$  & -5.1               & -5.1               \\
\enddata
\tablenotetext{a}{HD\,90043 \\
}
\end{deluxetable}

Atmospheric parameters of the target stars are estimated from
iodine-free, ``template'' spectra using the LTE
spectroscopic analysis package {\it Spectroscopy Made Easy}
\citep[SME;][]{valenti96}, as described by 
\citet{valenti05} and \citet{fischer05b}. To constrain the low surface
gravities of the evolved stars we used the iterative scheme of
\citet{valenti09}, which ties the SME-derived value of $\log{g}$ to
the gravity inferred from the Yonsei-Yale \citep[Y$^2$;][]{y2} stellar model
grids. The analysis yields a best-fitting 
estimate of \teff, \logg, \feh, and \vsini. The properties of our
targets from Lick and Keck are listed in the fourth 
edition of the 
Spectroscopic Properties of Cool Stars Catalog (SPOCS IV.; Johnson et
al. 2010, in prep). We adopt the SME parameter uncertainties
described in the error analysis of \citet{valenti05}. 

The luminosity of each star is estimated from the apparent V-band
magnitude and the
parallax from {\it Hipparcos} \citep{hipp2}, together with the
bolometric correction from \citet{vandenberg03}.  From \teff\ and
luminosity, we determine 
the stellar mass, radius, and an age estimate by associating those
observed properties with a model from the Y$^2$ stellar
interior calculations \citep{y2}. We also measure the chromospheric
emission in the CaII\,H\&K line cores 
\citep{wright04b}, providing an $S$ value on the
Mt. Wilson system, which we convert to \rphk\ as per \cite{Noyes84}.
The stellar properties of the host stars are summarized in Table
\ref{tab:stars}.  

\section{Keplerian Orbital Solutions}
\label{sec:orbits}

In this section we present the RV time series for both stars and the
initial orbital analysis, which consists of the sum of two
Keplerians without 
gravitational interaction. In \S~\ref{sec:dynamical} we present the
results of our Newtonian dynamical analysis for each system, which
properly accounts for non-Keplerian motion due to gravitational
interactions between the planets and host star.

\begin{deluxetable*}{llllll}[!ht]
\tablecaption{Orbital Parameters from Non-interacting Two-planet Solution\label{tab:planets}}
\tablewidth{0pt}
\tablehead{\colhead{Parameter} &
  \colhead{\starA\,b\tablenotemark{b}}    &
  \colhead{\starA\,c}    &
  \colhead{\starB\,b}    &
  \colhead{\starB\,b}    \\
}
\startdata
Period (d)    &
   \pAb~(\peAb)    &
   \pAc~(\peAc)    &
   \pBb~(\peBb)    &
   \pBc~(\peBc)    \\
T$_p$\tablenotemark{b}~(JD)    &
   \tpAb~(\tpeAb)    &
   \tpAc~(\tpeAc)    &
   \tpBb~(\tpeBb)    &
   \tpBc~(\tpeBc)    \\
Eccentricity     & 
   \eAb~(\eeAb)    &
   \eAc~(\eeAc)    &
   \eBb~(\eeBb)    &
   \eBc~(\eeBc)    \\
K~(\ms)   & 
   \kAb~(\keAb)    &
   \kAc~(\keAc)    &
   \kBb~(\keBb)    &
   \kBc~(\keBc)    \\
$\omega$~(deg) & 
   \omAb~(\omeAb)  &
   \omAc~(\omeAc)  &
   \omBb~(\omeBb)  &
   \omBc~(\omeBc)  \\
\msini~(\mjup) & 
   \msiniAb~(\msinieAb) &
   \msiniAc~(\msinieAc) &
   \msiniBb~(\msinieBb) &
   \msiniBc~(\msinieBc) \\
$a$~(AU)      &
   \arelAb~(\areleAb) &
   \arelAc~(\areleAc) &
   \arelBb~(\areleBb) &
   \arelBc~(\areleBc) \\
Lick rms~(\ms) & 
   \rmslA &
   ...   &
   \rmslB &
   ...   \\
Keck rms~(\ms) & 
   \rmskA &
   ...   &
   \rmskB &
   ...   \\
Jitter~(\ms)  & 
5.0 &
... &
5.0 &
... \\
\chisq        & 
    \chisA &
    ...   &
    \chisB &
    ... \\
N$_{\rm obs}$ Lick& 
    \nobslA &
    ... &
    \nobslB &
    ... \\
N$_{\rm obs}$ Keck& 
    \nobskA &
    ... &
    \nobskB &
    ... \\
\enddata
\tablenotetext{a}{HD\,90043}
\tablenotetext{b}{Time of periastron passage.}
\end{deluxetable*}

To search for the best-fitting, two-planet orbital
solution for each time series we use the partially-linearized
technique presented by \citet{wrighthoward}, as implemented in the
{\tt IDL} software suite {\tt RVLIN}. We estimate the parameter
uncertainties using a Markov-Chain Monte Carlo (MCMC) algorithm, as
described by \citet{bowler10}. 

\begin{figure}[!t]
\epsscale{1.1}
\plotone{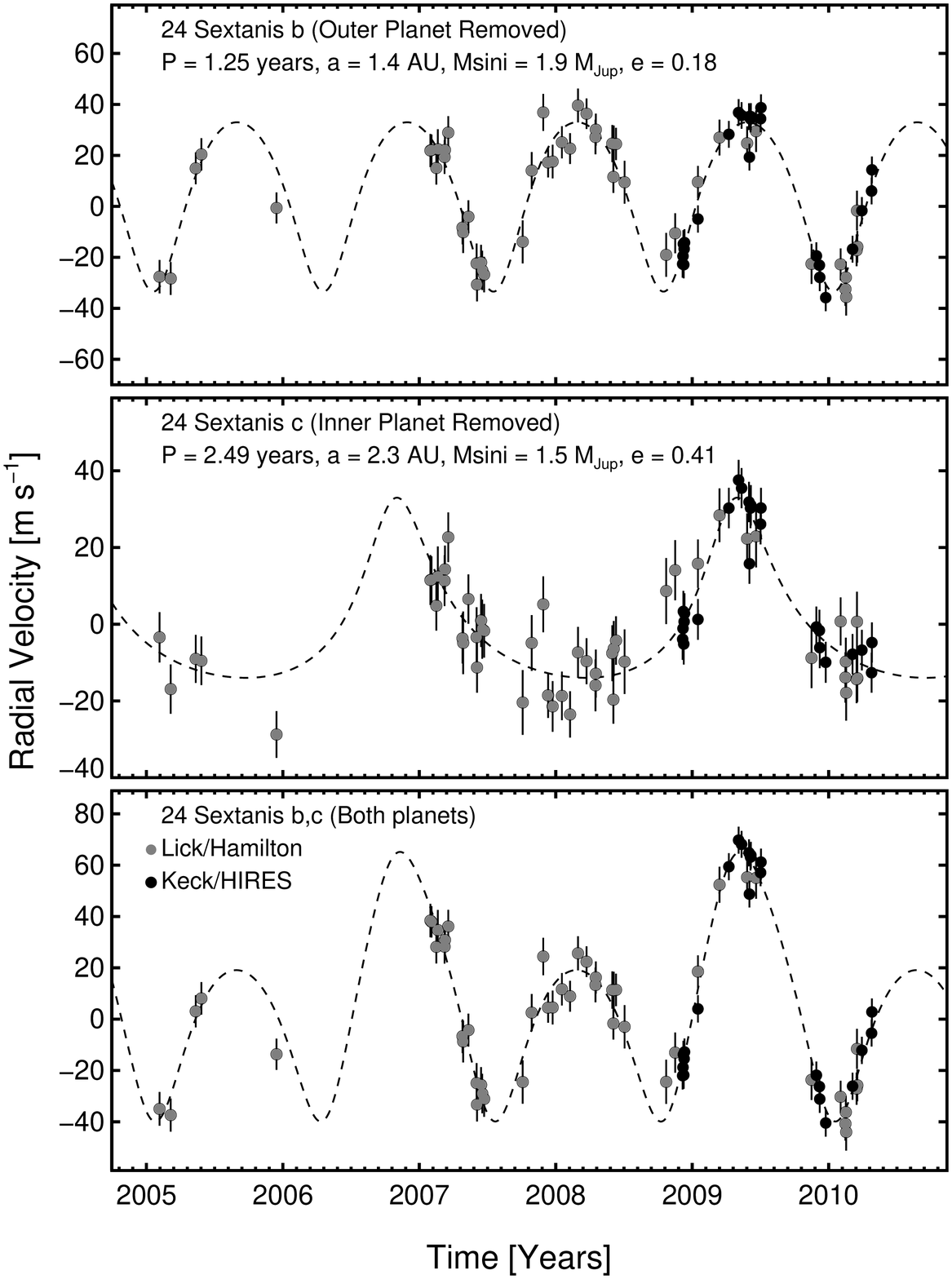}
\caption {RV time series and two-Keplerian orbital solution for
  \starA. {\it Top:} Inner 
  planet  with the signal from the outer planet 
  removed. The dashed line shows the best-fitting orbital solution for
  \starA\,b.
  {\it Middle:} The outer planet with the inner planet
  removed. The dashed line shows the best-fitting orbital solution for
  \starA\,c. {\it Bottom:} The full RV time series, with the two-planet
  solution shown as a dashed line. \label{fig:orbitA}}   
\end{figure}

\subsection{24~Sextanis}

We obtained initial-epoch observations of \starA\ at Lick Observatory
in 2005 February, and since then we have obtained \nobslA\ RV
measurements. After noticing time-correlated RV variations, we 
began additional monitoring at Keck Observatory in 2008
December, where we have obtained \nobskA\ additional RV measurements. The
RVs from both observatories are listed in Table~\ref{tab:rvA}, along
with the Julian Dates (JD) of observation and the internal measurement
uncertainties (without jitter). Fig.~\ref{fig:orbitA} shows the RV
time series from both observatories, where the error bars represent the
quadrature sum of the internal errors and 5~\ms\ of jitter. 

\citet{bowler10} reported evidence of a two-planet system around
\starA, but the data at the time could not provide a unique orbital
solution. An additional season of observations has provided stronger
constraints on the possible orbits of the two planets. We
fitted a model consisting of two non-interacting planets and a
\mstarA~\msun\ star orbiting their mutual center of mass. We find that
two Keplerians provide an acceptable fit to the data with an rms
scatter of 6.8~\ms\ and a reduced \chisq~$ = \chisA$. The inner planet
has a period of $P = \pAb \pm \peAb$~days, velocity semiamplitude $K =
\kAb \pm \keAb$~\ms, and eccentricity $e = \eAb \pm \eeAb$.  The outer
planet has $P = \pAc \pm \peAc$~days, $K = \kAc \pm \keAc$~\ms, and
$e = \eAc \pm \eeAc$. Together with our stellar mass estimate these
spectroscopic orbital parameters yield semimajor axes $\{a_b,a_c\} =
\{\arelAb,\arelAc\}$\,AU and minimum planet masses 
$\{M_b,M_c\}\sin{i} = \{\msiniAb,\msiniAc\}$~\mjup. 
A more detailed, dynamical analysis presented in
\S~\ref{sec:dynamical} revises this two-Keplerian solution under
the constraint of long-term stability. 

\begin{figure}[!t]
\epsscale{1.1}
\plotone{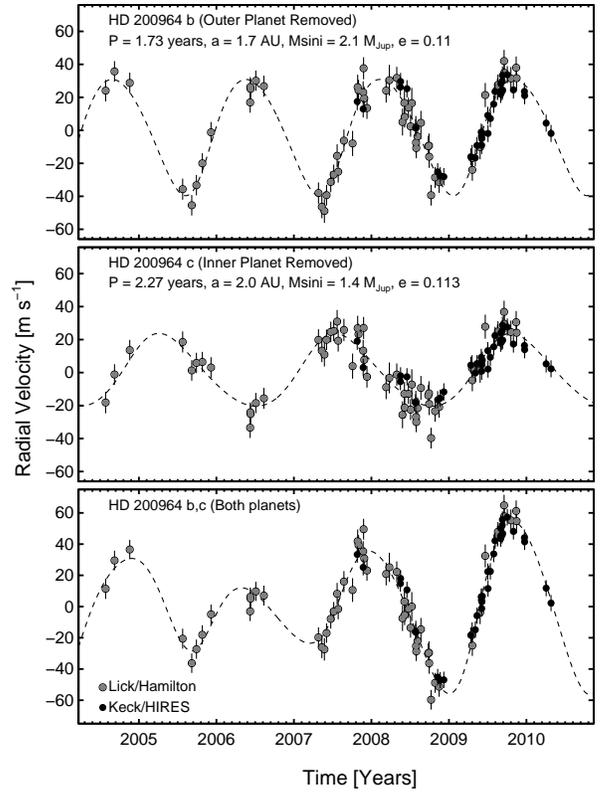}
\caption {RV time series and two-Keplerian orbital solution for
  \starB. {\it Top:} Inner 
  planet  with the signal from the outer planet 
  removed. The dashed line shows the best-fitting orbital solution for
  \starB\,b.
  {\it Middle:} The outer planet with the inner planet
  removed. The dashed line shows the best-fitting orbital solution for
  \starB\,c. {\it Bottom:} The full RV time series, with the two-planet
  solution shown as a dashed line. \label{fig:orbitB}}   
\end{figure}

\subsection{HD\,200964}

We began monitoring \starB\ at Lick Observatory in 2007~July and
have obtained \nobslB\ RV measurements. Time-correlated variations in
the star's RVs prompted additional monitoring at Keck Observatory
where we have obtained \nobskB\ measurements since 2007~October. 
The RVs from both observatories are listed in Table~\ref{tab:rvB},
along with the Julian Dates (JD) of observation and the internal
measurement uncertainties (without jitter). Fig.~\ref{fig:orbitB} shows
the RV time series from both observatories, where the error bars
represent the quadrature sum of the internal errors and 5~\ms\ of
jitter.  

As is the case for \starA, \citet{bowler10} reported evidence of a
two-planet system around 
\starB, and an additional season of observations has provided stronger
constraints on the possible orbits of the planets in the system. We
find that a two-Keplerian model provides an acceptable fit to the data
with an rms scatter of 6.8~\ms\ and a reduced \chisq~$ = \chisA$. The
inner planet 
has a period of $P = \pBb \pm \peBb$~days, velocity semiamplitude $K =
\kBb \pm \keBb$~\ms, and eccentricity $e = \eBb \pm \eeBb$.  The outer
planet has $P = \pBc \pm \peBc$~days, $K = \kBc \pm \keBc$~\ms, and
$e = \eBc \pm \eeBc$. Together with our stellar mass estimate these
spectroscopic orbital parameters yield semimajor axes $\{a_b,a_c\} =
\{\arelBb,\arelBc\}$\,AU and minimum planet masses 
$\{M_b,M_c\}\sin{i} = \{\msiniBb,\msiniBc\}$~\mjup. A more in-depth
dynamical analysis presented in \S~\ref{sec:dynamical} revises this
two-Keplerian solution under the constraint of long-term stability.

\section{Photometric Monitoring}

In addition to the radial velocities presented in \S~\ref{sec:orbits}
we also acquired photometric measurements of both \starA\ and \starB\
with the T12 0.80~m automatic photometric telescope (APT) at Fairborn 
Observatory.  The T12 APT and its two-channel photometer measure count 
rates simultaneously through Str\"omgren $b$ and $y$ filters.  T12 is 
essentially identical to the T8 0.80~m APT described in \citet{henry99}. 
Each program star ($P$) was observed differentially with respect to two 
nearby comparison stars ($C1$ and $C2$) (see Table~\ref{tab:phot}).  The differential 
magnitudes $P-C1$, $P-C2$, and $C2-C1$ were computed from each set of
differential measures.  All differential magnitudes with internal 
standard deviations greater than 0.01 mag were rejected to eliminate
observations taken under non-photometric conditions.  The surviving 
observations were corrected for differential extinction with nightly 
extinction coefficients and transformed to the Str\"omgren photometric 
system with yearly-mean transformation coefficients.  We averaged the 
$b$ and $y$ measurements of each star into a single $(b+y)/2$ ``passband'' 
(which we designate $by$ in Table~\ref{tab:phot}) to improve the
precision of the  
brightness measurements.  Typical precision of a single $(b+y)/2$ 
observation, as measured for pairs of constant stars, is 
$\sim$0.0015--0.0025 mag on good photometric nights. 

\citet{queloz01b} and \citet{paulson04} have demonstrated how rotational 
modulation in the visibility of star spots on active stars can result 
in periodic radial velocity variations that mimic the presence of
a planetary companion.  Thus, the precise APT brightness measurements are
valuable for distinguishing between activity-related radial velocity
changes and true reflex motion of a star caused by a planet.  

Photometric results for \starA\ and \starB\ are given in Table~\ref{tab:phot}.  
Columns 7--9 give the standard deviations of the $P-C1$, $P-C2$, and 
$C2-C1$ differential magnitudes in the $(b+y)/2$ passbands.  All of 
the standard deviations are small and within the range of measurement 
precision with the T12 APT.  We also performed periodogram analyses
on all the data sets and found no significant periodicity between 
0.03 and 100 days that might be the signature of stellar rotation.  Our 
data sets are not sufficiently long to test for variability on the four 
planetary periods from Table~\ref{tab:planets}, but we expect any such
variability to be  
very small.  The photometric stability of both \starA\ and \starB\ and 
the long-term coherency of the observed radial velocity variations
provide strong support for the existence of all four new planets. 

\section{Dynamical Interactions}
\label{sec:dynamical}

Our best-fitting double-Keplerian fits have two Jupiter-mass planets
with periods near the 2:1 commensurability for \starA\ and near 4:3
for \starB. The
proximity of the two planets implies strong gravitational
interactions, which limits the number of possible orbits to those that
allow the two planets to remain stable over the lifetime of the
star. To test the long-term stability of the 
various orbital configurations that are consistent with the data we  
performed a series of numerical integrations as described in the
following sections.

\subsection{Methodology for MCMC Analysis incorporating N-Body
Integrations}\label{SEC:NBODYMETHOD} 
We analyze the radial velocity observations using a Bayesian framework
following \citet{ford05} and \citet{ford06b}.  We assume priors that
are uniform in logarthimic intervals of orbital period, eccentricity,
argument of pericenter, mean 
anomaly at epoch, and the velocity zero-point.  For the velocity
amplitude ($K$) and stellar jitter ($\sigma_j$), we adopt a prior of
the form $p(x)=(x+x_o)^{-1}[log(1+x/x_o)]^{-1}$, with
$K_o=\sigma_{j,o}=1$m/s. For a discussion of priors, see
\citet{ford07}. The likelihood for 
radial velocity terms assumes that each radial velocity observation
($v_i$) is independent and normally distributed about the true radial
velocity with a variance of $\sigma_i^2+\sigma_j^2$, where $\sigma_i$
is the published measurement uncertainty and $\sigma_j$ is a jitter
parameter that accounts for additional scatter due to stellar
variability, instrumental errors and/or inaccuracies in the model
(i.e., neglecting planet-planet interactions or additional,
low-amplitude planet signals). 

In our initial phase of analysis, we use an MCMC method based upon
Keplerian orbit fitting to calculate a sample from the posterior
distribution \citep{ford06b}.  We
calculate multiple Markov chains, each with $\sim~2\times 10^8$
states.  We discard the first half of the chains and calculate
Gelman-Rubin test statistics for each model parameter and several
ancillary variables.  We find no indications of non-convergence. Thus,
we randomly choose a subsample ($\sim 16,000$ samples) from the
posterior distribution for further investigation. 

\begin{deluxetable*}{llllll}
\tablecaption{Orbital Parameters from Stable, Interacting N-Body Two-planet Solution\label{tab:nbodyplanets}}
\tablewidth{0pt}
\tablehead{\colhead{Parameter} &
  \colhead{\starA\,b\tablenotemark{b}}    &
  \colhead{\starA\,c}    &
  \colhead{\starB\,b}    &
  \colhead{\starB\,c}    \\
}
\startdata
Period (d)    &
   \NBODYpAb~(\NBODYpeAb)    &
   \NBODYpAc~(\NBODYpeAc)    &
   \NBODYpBb~(\NBODYpeBb)    &
   \NBODYpBc~(\NBODYpeBc)    \\
T$_p$\tablenotemark{b}~(JD)    &
   \NBODYtAb~(\NBODYteAb)    &
   \NBODYtAc~(\NBODYteAc)    &
   \NBODYtBb~(\NBODYteBb)    &
   \NBODYtBc~(\NBODYteBc)    \\
Eccentricity     & 
   \NBODYeAb~(\NBODYeeAb)    &
   \NBODYeAc~(\NBODYeeAc)    &
   \NBODYeBb~(\NBODYeeBb)    &
   \NBODYeBc~(\NBODYeeBc)    \\
K~(\ms)   & 
   \NBODYkAb~(\NBODYkeAb)    &
   \NBODYkAc~(\NBODYkeAc)    &
   \NBODYkBb~(\NBODYkeBb)    &
   \NBODYkBc~(\NBODYkeBc)    \\
$\omega$~(deg) & 
   \NBODYomAb~(\NBODYomeAb)  &
   \NBODYomAc~(\NBODYomeAc)  &
   \NBODYomBb~(\NBODYomeBb)  &
   \NBODYomBc~(\NBODYomeBc)  \\
\msini~(\mjup) & 
   \NBODYmsiniAb~(\NBODYmsinieAb) &
   \NBODYmsiniAc~(\NBODYmsinieAc) &
   \NBODYmsiniBb~(\NBODYmsinieBb) &
   \NBODYmsiniBc~(\NBODYmsinieBc) \\
$a$~(AU)      &
   \NBODYarelAb~(\NBODYareleAb) &
   \NBODYarelAc~(\NBODYareleAc) &
   \NBODYarelBb~(\NBODYareleBb) &
   \NBODYarelBc~(\NBODYareleBc) \\
Jitter~(\ms)  & 
   \NBODYjA~(\NBODYjeA) &&
   \NBODYjB~(\NBODYjeB) &\\
\enddata
%\tablenotetext{a}{HD\,90043}
%\tablenotetext{b}{Time of periastron passage.}
\end{deluxetable*}

Next, we use this subsample as the basis for a much more
computationally demanding analysis that uses fully self-consistent
n-body integrations to account for planet-planet interactions when
modeling the RV observations.  We again perform a Bayesian analysis,
but replace the standard MCMC algorithm with a Differential
Evolution Markov chain Monte Carlo (DEMCMC) algorithm 
\citep{terbraak06,veras09b, veras10}. In the DEMCMC algorithm each state of the
Markov chain is an ensemble of orbital solutions.  The candidate
transition probability function is based on the orbital parameters in
the current ensemble, allowing the DEMCMC algorithm to sample more
efficiently from high-dimensional parameter spaces that have strong
correlations between model parameters. More details of this DEMCMC
algorithm and associated tests of its accuracy will be presented in a
forthcoming paper, (Nelson et al., 2011, in prep.)

The priors for the model parameters are the same as those of the MCMC
simulations.  The initial conditions of the n-body simulations are
calculated by converting between Keplerian and Cartesian
coordinates. In this paper, we consider only coplanar
two-planet systems. 

For the n-body integrations, we use a time symmetric 4th order Hermite
integrator that has been optimized for planetary systems (Kokubo et
al.\ 1998).  We extract the radial velocity of the star (in the
barycentric frame) at each of the observation times for comparison to
RV data.  During the DEMCMC analysis, we also impose the constraint of
short-term (100 years) orbital stability.  We check whether the
planetary semimajor axes remain within a factor of 50\% of their
starting value, and that no close-approaches occur within $0.1\times$
the semimajor axis during the 100 year n-body integration.  Any
systems failing these tests are rejected as unstable (regardless of
the quality of the fit to RV data).  Thus, the DEMCMC simulations
avoid orbital solutions that are violently unstable.  In our DEMCMC
simulations, this process is repeated for 10,000 generations, each of
which contains 16,000 systems, for a total of $\sim 10^8$ n-body
integrations in each DEMCMC simulation. 

Since the DEMCMC simulations only require stability for $\sim~100$
years, the orbital solutions in the final generation may or may not be
stable for longer timescales.  Since nearly all of these systems are
strongly interacting, we take this final generation (16k systems) and
demand that they also be stable (according to the same criteria above)
over the course of a $10^7$ year integration, performed using the
hybrid Bulirsch-Stoer / Symplectic  integrator {\sc Mercury}
\citep{chambers99}. Only the orbital solutions
which are stable over the 
course of this long-term n-body integration are regarded as being
acceptable solutions. More details on the dynamical analysis performed
and the results obtained will be presented in a companion paper,
(Payne et al, 2010, in prep.).

\subsection{Numerical Integrations for \starA}
\begin{figure}[!t]
\epsscale{1.1}
\plotone{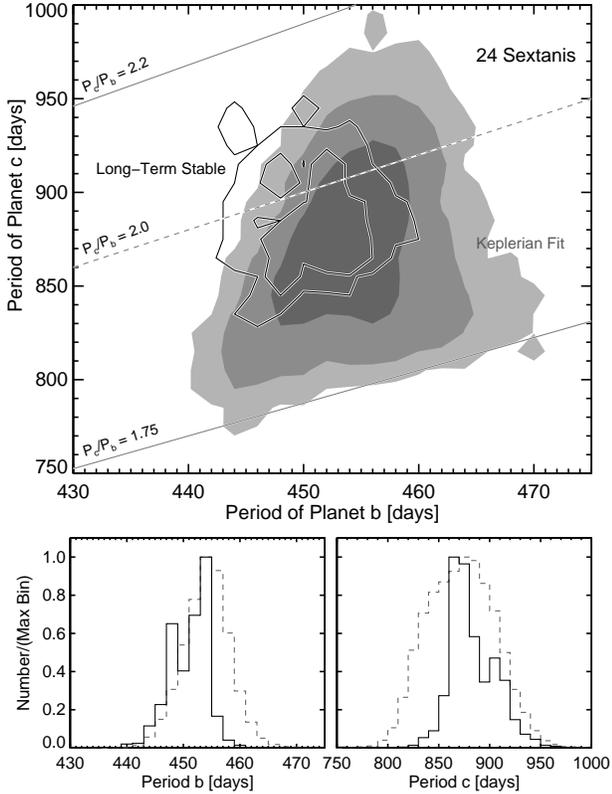}
\caption{Top: Comparison of the end-points of the
  double-Keplerian MCMC analysis (shaded contours), and the n-body
  DEMCMC analysis with the long-term stable constraint
  (black, solid contours) for \starA. The contours show the 68.2\%,
  80\% and 95\% confidence intervals (inner to outer). To guide the eye, we
  overlay diagonal lines to indicate 1.75:1, 2:1 \& 2.2:1 period ratios. The
  MCMC posterior sample covers a large range of parameter space and is
  consistent with a 2:1 period ratio. In contrast, the DEMCMC results
  are limited to a smaller area of parameter space lying on, or close
  to, the 2:1 period ratio. Bottom: Histograms for the same planetary period
  data from \starA, comparing the results from the Keplerian MCMC
  analysis (black), N-Body DEMCMC analysis (dark gray). 
  When compared to the basic MCMC results, the DEMCMC
  results for the inner planet are shifted more towards lower
  periods, while the outer planet shifts towards higher
  periods, making the overall period ratio converge more closely
  towards 2:1.} 
  \label{FIG:NBODY:A:1}
\end{figure}

We find that the n-body DEMCMC routine results for \starA\ are
concentrated around solutions with $\{P_b,P_c\} \approx \{450,900\}$ days,
i.e. close to the 2:1 resonance, and that they span a significantly
smaller range of parameter space than do the double-Keplerian
fits. The n-body RV fitting aspect of the DEMCMC routine acts
to shrink the parameter space, whereas the stability requirement in
the routine has the effect of shifting the DEMCMC solutions towards the 2:1
period ratio mark, primarily by favoring lower periods for the inner
planet. We illustrate in Fig. \ref{FIG:NBODY:A:1} the difference in
the end-points of the two analyses, showing a scatter plot of the
planetary periods at the end of both the Keplerian analysis and the
n-body analysis. 

We have applied the DEMCMC method described in
\S~\ref{SEC:NBODYMETHOD} multiple times using different values for the
initial ensemble of orbital solutions.  Each time, these simulations
reached qualitatively similar results to those of
Fig. \ref{FIG:NBODY:A:1}, but quantitatively there are signs that the
method has not fully converged.  Therefore, we do not interpret the
results as a precise estimate of the posterior probability
distribution.  Instead, we use these results to demonstrate that there
exist orbital solutions that are both stable and consistent with the
Doppler observations. Most importantly, all our simulations indicate
that the posterior distribution is concentrated at solutions with a
ratio of orbital periods between $\sim 1.9$ and $\sim 2.1$ and very
close to the 2:1 MMR. Thus, we conclude that the current observations
strongly favor orbital solutions in (or at least near) the 2:1 MMR. 

Next, we performed long-term stability tests for the 16,000 orbital
solutions in the final generation of the n-body DEMCMC analysis.  We
find that $\sim 90\%$ of the orbital solutions are unstable over the
course of $10^7$ years of integration.  We consider the remaining
$\sim 1,000$ (of the 16,000 orbital solutions) that are stable for
$10^7$ years to be plausible orbital solutions given both the RV data
for \starA\ and the requirement of long-term dynamical stability. 

We show in Fig. \ref{FIG:NBODY:A:1} that the stable systems remain
strongly clustered around the 2:1 period commensurability
region. Taking all of these long-term stable systems into account, we
find that the data indicate that the inner planet has a period $P_b =$
\NBODYpAb~\NBODYpeAb~days, semimajor axis $a_b =$ \NBODYarelAb
\NBODYareleAb~AU, eccentricity $e_b =$ \NBODYeAb \NBODYeeAb\ and mass
$M_b \sin{i} =$ \NBODYmsiniAb \NBODYmsinieAb~\mjup, while the outer
planet has a period $P_c =$ \NBODYpAc \NBODYpeAc~days, semimajor axis
$a_c =$ \NBODYarelAc \NBODYareleAc~AU, eccentricity $e_c =$ \NBODYeAc
\NBODYeeAc\ and mass $M_c \sin{i} =$ \NBODYmsiniAc
\NBODYmsinieAc~\mjup. We detail all the fitted parameters from this
analysis in Table~\ref{tab:nbodyplanets}. 

When we examine specific systems in detail, we find that the
pericenter of the outer planet overlaps the location of the
pericenter of the inner planet, meaning that the planets therefore
experience detectable gravitational 
interactions. The systems typically undergo large eccentricity
oscillations ($0.24 < e_b < 0.47$, $0.5 < e_c < 0.6$) over a period of
$\sim 800$ years, while the semimajor axis varies on a shorter ($\sim
25$ year) timescale, with a range $1.29 < a_b < 1.31$ and
$2.04 < a_c < 2.10$.  

We note that the PDFs from the DEMCMC analysis are bimodal.  To
investigate the cause of the biomodality, we performed a similar
n-body$+$DEMCMC analysis without the requirement of short-term
stability.  The resulting PDFs have a single broader mode consistent
with the output of Keplerian MCMC analysis. We conclude that the
biomodal nature of the PDF from the dynamical analysis is most likely
the result of demanding short-term stability.

The overlap of the pericenters suggests that a MMR is
needed to stabilize the system over long timescales by preventing close
encounters. However, an analysis of the resonant angles
$\theta_{2,1}=2\lambda_c-1\lambda_b-1\varpi_{2,1}$ (where $\lambda$ is
the mean longitude and $\varpi$ is the longitude of pericenter)
suggests that, for all the long-term-stable systems, the planets circulate,
rather than librate, i.e. all of the systems are observed to have angular
ranges for $\theta_{2,1}\sim 360^{\circ}$. We therefore caution that
the current state of the observations and dynamical analysis cannot
confirm whether the system is truly in a resonant state, and hence
further work will be required. A more
detailed investigation (Payne et al. 2010 in prep) will probe
the nature of these dynamical interactions in \starA\ and \starB\
in greater detail.  

Finally, we note that as discussed in \S~\ref{SEC:NBODYMETHOD}, jitter
is allowed to vary during the DEMCMC procedure. As such we find a best
fit value for the jitter in the same manner as we do for the various
planetary orbital parameters. From the \starA\ analysis, we find a
large jitter value of \NBODYjA \NBODYjeA \ms. This is substantially
larger than the value of 5~\ms assumed in \S~\ref{sec:orbits},
indicating that \starA\ has intrinsic RV variability at the high end
of the observed distribution, or that there are other unmodeled
sources of variability such as an additional planet in the
system. However, a periodogram of the residuals shows no significant
power above the noise. Using the approach of \citet{bowler10}, we can
rule out additional companions with masses greater than 0.3~\mjup\ out
to 1~AU and ``hot'' planets with masses \msini~$< 0.1$~\mjup\ within
0.1~AU. Additional monitoring at Keck and a more in-depth dynamical 
analysis will help clarify the situation.

\begin{figure}[!t]
\epsscale{1.1}
\plotone{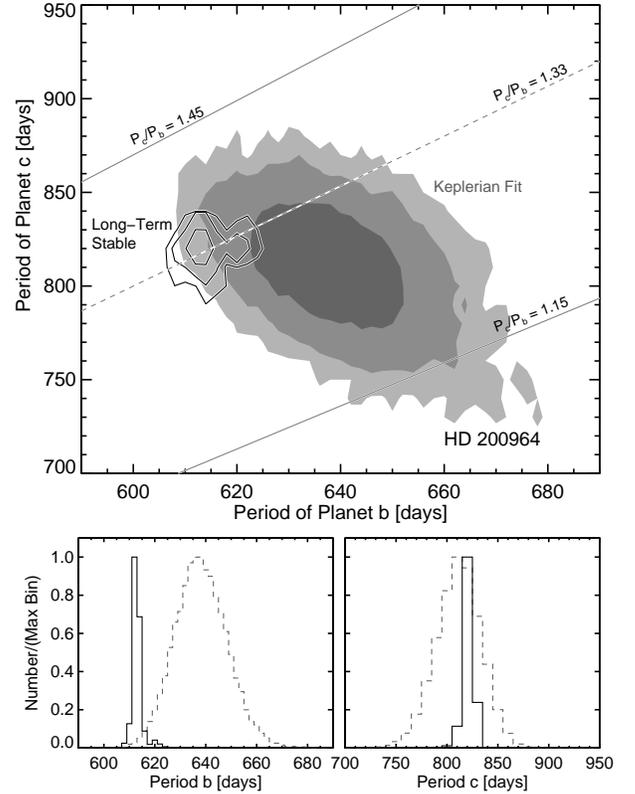}
\caption{Top: Comparison of the end-points of the
  double-Keplerian MCMC analysis (shaded contours), and the n-body
  DEMCMC analysis with the long-term stable constraint
  (black, solid contours) for \starA. The contours show the 68.2\%,
  80\% and 95\% confidence intervals (inner to outer). To guide the eye, we
  overlay diagonal lines to indicate 1.15:1, 4:3 \& 1.45:1 period
  ratios. The 
  MCMC posterior sample covers a large range of parameter space and is
  consistent with a 4:3 period ratio. In contrast, the DEMCMC results
  are limited to a smaller area of parameter space straddling 
  the 4:3 period ratio. Bottom: Histograms
  for the two planetary periods in the HD 200964 system, comparing the
  results from the Keplerian MCMC analysis (black) and N-Body
  DEMCMC analysis (dark gray). When compared to the basic
  MCMC results, the DEMCMC results in a decrease in the fitted
  period of the inner planet, making the overall period ratio
  converge more closely towards 4:3. \label{FIG:NBODY:B:1}} 
\end{figure}

\subsection{Numerical Integrations for \starB}
The results of our n-body DEMCMC simulations for 
\starB\ are again
more tightly confined than the results of the double-Keplerian
analysis. As an example of this we illustrate in
Fig. \ref{FIG:NBODY:B:1} the difference in the end-points of the two
analyses, showing a scatter plot of the planetary periods at the end
of both the Keplerian analysis and the n-body analysis, as well as
histograms for the same data. Again, we note that the shift to an
N-Body analysis acts to concentrate the preferred region into a
smaller area, while the stability requirements act to shift the
favored parameters closer toward the 4:3 period commensurability. 

As for the case of \starA, we have applied our DEMCMC method
multiple times using numerous randomized initial conditions. Again, we
find qualitatively similar results from all sets of the DEMCMC
analysis, but with some indication that the results have not
truly converged. As such, we interpret these results as
demonstrative that stable systems exist which can fit the RV data,
and moreover, the extremely narrow range of period ratios
favored by the analysis ($\sim 1.32$ to $\sim 1.36 $) shows that the
current observations strongly favor orbital solutions in or close to
the 4:3 MMR. 

Next, we test the long-term orbital stability of the orbital solutions
identified by the DEMCMC analysis.  For \starB, $> 90\%$ of the
systems are clearly unstable during a $10^7$ year integration (i.e.,
they experience a collision or a change in semimajor axes of more
than 50\%.)  We find that $\sim~1,000$ of the 16,000 simulations
remain stable for $10^7$ years.  We interpret these as plausible
orbital solutions consistent with the RV observations of HD 200964. 

As illustrated in Fig. \ref{FIG:NBODY:B:1}, the stable systems
occupy a region of parameter space corresponding to a region of
paramter space near the 4:3 period
ratio. Taking only these
long-term stable systems into account, we find that the inner planet
has a period $P_b =$ \NBODYpBb \NBODYpeBb~days, semimajor axis $a_b
=$ \NBODYarelBb \NBODYareleBb~AU, eccentricity $e_b =$ \NBODYeBb
\NBODYeeBb\ and mass $M_b \sin{i} =$ \NBODYmsiniBb \NBODYmsinieBb~\mjup, while the outer planet has a period $P_c =$ \NBODYpBc
\NBODYpeBc\ days, semimajor axis $a_c =$ \NBODYarelBc
\NBODYareleBc~AU, eccentricity $e_c =$ \NBODYeBc \NBODYeeBc\ and mass
$M_c \sin{i} =$ \NBODYmsiniBc \NBODYmsinieBc~\mjup. We detail all
the fitted parameters from this analysis in Table
\ref{tab:nbodyplanets}.   

We 
find that most of the stable planetary orbits overlap, producing
strongly interacting 
systems, resulting in significant oscillations in the the
semimajor axes and the eccentricities of both
planets. An example stable solution exhibits observable eccentricity
oscillations ($ 0.03 < e_b < 
0.1$ and $ 0.13 < e_c < 0.18$) on an approximately 250-year timescale
and semimajor axis oscillations ($ 1.57 < a_b < 1.58$ and $ 1.9 < a_c
< 1.93$) on an approximately 60 year timescale. Again, as in the
\starA\ system, all of the long-term stable systems examined here
for \starB\ seem to be circulating rather librating.
As in the \starA\ analysis, we find a jitter value of \NBODYjB
\NBODYjeB \ms, which is larger than the empirical jitter 
estimate of \citet{johnson10b}.

\section{Summary and Discussion}

Our RV measurements of the intermediate-mass subgiants 24~Sex and
HD\,200964 ($M_\star = $~\mstarA\ and \mstarB~\msun, respectively)
reveal the presence of a pair of giant planets around each star. Our
orbital analysis indicates that most, if not all, of the dynamically
stable solutions include crossing 
orbits, suggesting that each system is locked in a mean motion
resonance that prevents close encounters and provides long-term
stability.  The planets in the \starA\ system likely have a period
ratio  near 2:1, while the \starB\ system is even more tightly
packed with a period ratio close to 4:3. 

In both the \starA\ and \starB\ systems, the planets reside well
within the so-called snow line, beyond which volatiles in the
protoplanetary disk can condense to provide the raw materials for
protoplanetary core growth. For a pre-main-sequence, 1.5~\msun\ star
the snow line is located beyond 2-3 AU according to the estimates of
\citet{kennedy08} for realistic disk models including irradiation and
accretional heating. It is therefore likely that the planets around
\starA\ and \starB\ formed at 
larger semimajor axes and subsequently experienced inward orbital
migration \citep[See][for a review of migration theory]{migrationrev}. 

Unless the planets around \starA\ and \starB\ formed with period
ratios close to their current 
values, the two planets in each system likely migrated through disk
interactions at convergent rates until they became trapped in an MMR,
with the strong 2:1 resonance being the most likely
endpoint of such differential migration \citep{kley00, nelson02}.  The
rarity of planets 
discovered with period ratios smaller than 2:1 accords well with the
dynamical simulations of \citet{lee09}. In their
simulation they considered the formation of two giant planets in a
protoplanetary disk with initial period ratios just outside of 2:1 and
final orbital configurations determined by the initial conditions and
details of the planet-planet-disk interactions. From their ensemble of
simulated planetary systems they found that only 3\% attain period
ratios closer than 2:1, and none ended in stable configurations
closer than 3:2.  

The ability of planets to reach a resonance tighter than 3:2 is
dependent upon a number of factors including the initial separation of
the planets, disk viscosity, planet masses, 
remaining disk mass and size of the gaps opened by the
planets \citep{malhotra93,nader99,bryden00,snellgrove01}. One key
variable is the convergent migration rate, 
which if fast enough can carry the planets past the deep 2:1 MMR into
tighter commensurabilities. The 24\,Sex system is near the 2:1 resonance,
which may be a reflection of a slower migration process compared to
that which led to the extremely tight 4:3 configuration in the
otherwise very similar HD\,200964 system.   

\citet{pap10} explored rapid migration scenarios leading to the
attainment of high-order MMRs by low-mass planets migrating withing a
gas disk. For the terrestrial planets they considered, MMRs with
degrees as high as 8:7 and 11:10 were achieved for migration
timescales of order $10^3$ years. However, to test whether such
conditions can lead to high-order stable MMRs for the planets in
the \starB\ system, and the 2:1 MMR seen in the \starA\ system,
hydrodynamic considerations need to be incorporated into the
models. \citet{rein10} investigated the formation and evolution of the
gas giants orbiting HD\,45364 and found plausible models for the
attainment of the 3:2 MMR observed in that system. Simulations of this
nature are beyond the scope of the present 
work and will be presented in a future contribution (Payne et
al. 2010, in prep). For now, it is clear that just like the resonant
planetary systems discovered before them, the \starA\ and
\starB\ systems pose interesting challenges to theories of planet
formation and orbital evolution. 

\acknowledgments

We thank the many observers who contributed to the observations
reported here. We gratefully acknowledge the efforts and dedication
of the Keck Observatory staff, especially Grant Hill, Scott Dahm and
Hien Tran for their support of 
HIRES and Greg Wirth for support of remote observing. We are also
grateful to the time assignment committees of NASA, NOAO, Caltech, and
the University of California for their generous allocations of
observing time. A.\,W.\,H.\ gratefully acknowledges support from a
Townes Post-doctoral Fellowship at the U.\,C.\ Berkeley Space Sciences 
Laboratory. E.\,B.\,F and M.\,J.\,P were supported by NASA Origins of 
Solar Systems grant NNX09AB35G. G.\,W.\,M.\ acknowledges NASA grant
NNX06AH52G.    G.\,W.\,H acknowledges support from NASA, NSF,
Tennessee State University, and the State of Tennessee through its
Centers of Excellence program. 
Finally, the authors wish to extend special thanks to those of
Hawaiian ancestry  on whose sacred mountain of Mauna Kea we are
privileged to be guests.   
Without their generous hospitality, the Keck observations presented herein
would not have been possible.

\bibliography{}

\clearpage

\begin{deluxetable}{llll}
\tablecaption{Radial Velocities for 24 Sextanis\label{tab:rvA}}
\tablewidth{0pt}
\tablehead{
\colhead{JD} &
\colhead{RV} &
\colhead{Uncertainty} &
\colhead{Telescope} \\
\colhead{-2440000} &
\colhead{(m~s$^{-1}$)} &
\colhead{(m~s$^{-1}$)} &
\colhead{}
}
\startdata
13405.936 &  -34.98 &  4.24 & L \\
13435.836 &  -37.40 &  4.05 & L \\
13502.774 &    3.04 &  3.76 & L \\
13517.768 &    8.07 &  3.90 & L \\
13719.011 &  -13.68 &  3.46 & L \\
14131.051 &   38.40 &  4.17 & L \\
14133.931 &   37.92 &  3.49 & L \\
14146.787 &   28.17 &  4.15 & L \\
14150.947 &   34.70 &  6.03 & L \\
14168.823 &   28.25 &  4.33 & L \\
14169.860 &   30.86 &  3.65 & L \\
14178.699 &   36.12 &  4.10 & L \\
14216.839 &   -6.66 &  4.22 & L \\
14218.709 &   -8.74 &  6.44 & L \\
14232.713 &   -4.29 &  4.05 & L \\
14254.688 &  -24.93 &  6.01 & L \\
14255.689 &  -33.28 &  4.29 & L \\
14266.699 &  -25.63 &  4.77 & L \\
14269.695 &  -28.99 &  5.87 & L \\
14274.719 &  -31.09 &  4.86 & L \\
14378.045 &  -24.52 &  6.79 & L \\
14402.054 &    2.53 &  5.26 & L \\
14433.046 &   24.41 &  5.29 & L \\
14445.938 &    4.46 &  3.08 & L \\
14458.050 &    4.53 &  4.31 & L \\
14482.959 &   11.65 &  3.89 & L \\
14504.855 &    8.92 &  3.36 & L \\
14525.946 &   25.65 &  4.37 & L \\
14548.900 &   22.38 &  3.29 & L \\
14572.872 &   13.31 &  4.54 & L \\
14573.724 &   16.22 &  3.77 & L \\
14617.764 &   11.34 &  5.29 & L \\
14620.705 &   -1.69 &  3.78 & L \\
14621.716 &   11.26 &  5.22 & L \\
14627.708 &   11.39 &  3.87 & L \\
14650.692 &   -2.99 &  6.79 & L \\
14762.024 &  -24.44 &  7.00 & L \\
14786.078 &  -13.05 &  6.03 & L \\
14806.150 &  -22.03 &  1.98 & K \\
14807.075 &  -18.78 &  1.83 & K \\
14808.057 &  -13.83 &  1.96 & K \\
14809.069 &  -21.81 &  1.88 & K \\
14810.149 &  -15.43 &  1.88 & K \\
14811.122 &  -12.80 &  1.65 & K \\
14847.038 &    3.99 &  1.80 & K \\
14847.048 &   18.53 &  3.83 & L \\
14904.830 &   52.37 &  4.91 & L \\
14929.814 &   59.40 &  1.63 & K \\
14955.878 &   69.74 &  1.60 & K \\
14963.915 &   68.12 &  1.49 & K \\
14978.706 &   55.29 &  4.31 & L \\
14983.770 &   64.81 &  1.59 & K \\
14984.827 &   48.72 &  1.53 & K \\
14987.759 &   63.15 &  1.54 & K \\
14988.750 &   63.85 &  1.60 & K \\
15002.711 &   55.03 &  6.40 & L \\
15014.741 &   57.09 &  1.72 & K \\
15015.750 &   61.18 &  1.58 & K \\
15151.052 &  -23.66 &  6.04 & L \\
15164.112 &  -21.86 &  1.83 & K \\
15172.150 &  -26.29 &  2.05 & K \\
15173.097 &  -31.15 &  1.93 & K \\
15189.119 &  -40.47 &  1.85 & K \\
15229.007 &  -30.27 &  3.76 & L \\
15241.871 &  -40.78 &  4.23 & L \\
15242.872 &  -36.26 &  3.75 & L \\
15243.855 &  -43.99 &  5.27 & L \\
15260.941 &  -26.13 &  1.65 & K \\
15271.797 &  -27.09 &  3.77 & L \\
15272.789 &  -11.58 &  6.00 & L \\
15273.750 &  -25.81 &  3.66 & L \\
15285.854 &  -12.17 &  1.75 & K \\
15311.810 &   -5.48 &  1.43 & K \\
15312.791 &    2.81 &  1.56 & K
\enddata
\end{deluxetable}

\begin{deluxetable}{llll}
\tablecaption{Radial Velocities for HD\,200964\label{tab:rvB}}
\tablewidth{0pt}
\tablehead{
\colhead{JD} &
\colhead{RV} &
\colhead{Uncertainty} &
\colhead{Telescope} \\
\colhead{-2440000} &
\colhead{(m~s$^{-1}$)} &
\colhead{(m~s$^{-1}$)} &
\colhead{}
}
\startdata
13213.895 &   11.40 &  4.20 & L \\
13255.775 &   29.55 &  3.72 & L \\
13327.604 &   36.56 &  3.70 & L \\
13576.944 &  -20.58 &  3.98 & L \\
13619.810 &  -36.28 &  3.71 & L \\
13641.795 &  -27.30 &  3.61 & L \\
13669.629 &  -18.07 &  3.55 & L \\
13710.605 &   -4.99 &  3.77 & L \\
13894.977 &   -3.08 &  3.71 & L \\
13895.921 &    6.24 &  3.30 & L \\
13896.963 &    5.08 &  3.47 & L \\
13921.953 &    9.72 &  3.59 & L \\
13959.788 &    7.02 &  3.96 & L \\
14216.976 &  -19.74 &  4.09 & L \\
14232.936 &  -26.06 &  4.83 & L \\
14244.907 &  -27.39 &  5.19 & L \\
14254.960 &  -16.97 &  3.70 & L \\
14274.916 &   -7.88 &  3.72 & L \\
14288.888 &   -3.15 &  3.49 & L \\
14304.877 &    8.18 &  4.89 & L \\
14309.845 &   -1.57 &  4.28 & L \\
14336.881 &   16.00 &  4.49 & L \\
14377.773 &   10.58 &  5.74 & L \\
14399.752 &   33.36 &  1.62 & K \\
14401.749 &   41.83 &  5.67 & L \\
14405.700 &   39.39 &  4.02 & L \\
14427.588 &   35.30 &  4.29 & L \\
14427.757 &   25.14 &  1.18 & K \\
14429.621 &   49.55 &  4.41 & L \\
14432.653 &   30.85 &  4.89 & L \\
14445.624 &   23.13 &  4.22 & L \\
14536.053 &   20.85 &  5.26 & L \\
14551.019 &   25.15 &  7.90 & L \\
14585.990 &   22.22 &  4.51 & L \\
14603.126 &   14.71 &  1.39 & K \\
14604.012 &   18.01 &  1.58 & K \\
14612.988 &   -7.49 &  6.44 & L \\
14621.965 &    3.21 &  5.14 & L \\
14622.995 &   -5.37 &  4.16 & L \\
14634.079 &   10.67 &  1.52 & K \\
14640.923 &   -1.39 &  3.34 & L \\
14650.960 &  -13.59 &  3.79 & L \\
14656.902 &    0.00 &  7.81 & L \\
14674.916 &  -16.16 &  1.39 & K \\
14675.850 &  -16.06 &  4.87 & L \\
14676.879 &  -25.35 &  4.49 & L \\
14677.922 &  -28.67 &  3.69 & L \\
14683.850 &  -22.03 &  4.24 & L \\
14699.804 &  -14.57 &  5.28 & L \\
14734.713 &  -29.88 &  3.82 & L \\
14737.790 &  -29.53 &  5.34 & L \\
14738.751 &  -36.28 &  4.54 & L \\
14747.711 &  -59.70 &  3.80 & L \\
14766.659 &  -48.87 &  3.96 & L \\
14778.803 &  -45.21 &  1.61 & K \\
14785.605 &  -51.33 &  4.06 & L \\
14790.734 &  -47.17 &  1.34 & K \\
14807.789 &  -46.94 &  1.60 & K \\
14935.136 &  -18.39 &  1.16 & K \\
14941.012 &  -18.06 &  6.19 & L \\
14941.989 &  -24.98 &  4.38 & L \\
14956.097 &  -14.88 &  1.50 & K \\
14964.120 &   -5.85 &  1.33 & K \\
14978.935 &   -3.81 &  5.48 & L \\
14984.070 &    6.28 &  1.43 & K \\
14985.095 &    4.20 &  1.62 & K \\
14986.112 &   -1.27 &  1.71 & K \\
14987.129 &    3.06 &  1.39 & K \\
14989.069 &    6.64 &  1.33 & K \\
15002.947 &   32.46 &  5.34 & L \\
15014.972 &   22.36 &  1.54 & K \\
15015.957 &   11.56 &  1.44 & K \\
15027.007 &   22.52 &  1.55 & K \\
15042.973 &   33.72 &  1.69 & K \\
15049.001 &   42.16 &  1.62 & K \\
15060.816 &   45.09 &  4.29 & L \\
15061.819 &   45.31 &  3.69 & L \\
15062.829 &   47.93 &  3.69 & L \\
15075.078 &   45.22 &  1.68 & K \\
15076.067 &   43.59 &  1.75 & K \\
15077.056 &   49.98 &  1.57 & K \\
15082.047 &   45.65 &  1.53 & K \\
15083.053 &   52.10 &  1.63 & K \\
15084.028 &   55.68 &  1.57 & K \\
15085.004 &   46.81 &  1.60 & K \\
15091.772 &   64.86 &  4.56 & L \\
15092.725 &   57.80 &  3.96 & L \\
15106.911 &   57.25 &  1.51 & K \\
15123.798 &   55.22 &  4.81 & L \\
15135.755 &   48.13 &  1.43 & K \\
15148.734 &   61.20 &  4.57 & L \\
15150.620 &   54.74 &  5.97 & L \\
15187.695 &   44.10 &  1.66 & K \\
15188.688 &   41.49 &  1.69 & K \\
15290.146 &   11.79 &  1.39 & K \\
15313.137 &    2.22 &  1.33 & K
\enddata
\end{deluxetable}

\begin{deluxetable}{lllccccccc}
\tablewidth{0pt}
\tablecaption{Summary of Photometric Observations From Fairborn Observatory\label{tab:phot}}
\tablehead{
\colhead{Program} & \colhead{Comparison} & \colhead{Comparison} & 
\colhead{Date Range} & \colhead{Duration} & \colhead{} & 
\colhead{$\sigma{(P-C1)}_{by}$} & \colhead{$\sigma{(P-C2)}_{by}$} &
\colhead{$\sigma{(C2-C1)}_{by}$} & \colhead{} \\
\colhead{Star} & \colhead{Star 1} & \colhead{Star 2} &
\colhead{(HJD $-$ 2,400,000)} & \colhead{(days)} & \colhead{$N_{obs}$} & 
\colhead{(mag)} & \colhead{(mag)} & \colhead{(mag)} &
\colhead{Variability} \\
\colhead{(1)} & \colhead{(2)} & \colhead{(3)} & \colhead{(4)} & 
\colhead{(5)} & \colhead{(6)} & \colhead{(7)} & \colhead{(8)} & 
\colhead{(9)} & \colhead{(10)}
}
\startdata
 \starA\ &  HD 87974 &  HD 89734 & 54438--55333 & 895 & 243 & 0.0022 & 0.0020 & 0.0020 & Constant \\
\starB\ & HD 201507 & HD 201982 & 54578--55167 & 589 & 100 & 0.0015 & 0.0014 & 0.0015 & Constant \\
\enddata
\tabletypesize{\scriptsize}
\end{deluxetable}

\end{document}